\documentclass[fleqn,usenatbib]{mnras}

\usepackage{newtxtext,newtxmath}

\usepackage[T1]{fontenc}

\DeclareRobustCommand{\VAN}[3]{#2}
\let\VANthebibliography\thebibliography
\def\thebibliography{\DeclareRobustCommand{\VAN}[3]{##3}\VANthebibliography}

\usepackage{graphicx}	
\usepackage{amsmath}	

\title[The contact frontier in a structured Galaxy]{The contact frontier in Galactic civilization networks: radial structure and the Galactic habitable zone}

\author[K. Tsumura]{
Kohji Tsumura$^{1,2}$\thanks{E-mail: tsumura@astr.tohoku.ac.jp}
\\
$^{1}$Frontier Research Institute for Interdisciplinary Science, Tohoku University, Sendai, Japan\\
$^{2}$Department of Natural Science, Faculty of Science and Engineering, Tokyo City University, Tokyo, Japan
}

\date{Accepted XXX. Received YYY; in original form ZZZ}

\pubyear{2026}

\usepackage{xcolor}

\begin{document}
\label{firstpage}
\pagerange{\pageref{firstpage}--\pageref{lastpage}}
\maketitle

\begin{abstract}
The probability of interstellar contact depends not only on the abundance of technological civilizations but also on their spatial connectivity.
Treating civilizations as nodes linked whenever their separation is smaller than the distance a signal travels within a communicative lifetime, 
we determine the radial structure of Galactic civilization networks.
An exponential stellar disc replaces the uniform disc of earlier work, 
and the fraction of stars hosting civilizations may vary with radius as a Galactic Habitable Zone (GHZ).
The isolation, contact and network transitions are frontiers several kiloparsecs wide, not boundaries: 
the contact frontier is set entirely by the disc scale length, and the network frontier, broadened by gradient percolation, stays near 1.5 scale lengths.
The two frontiers overlap over a band containing the Sun, where both transitions proceed at once.
Within this factorized GHZ model, the radius of peak civilization density depends only logarithmically on the strength of supernova sterilization, 
and the wide range of published peak radii, from the Galactic centre to 12 kpc, is reproduced primarily by varying that one parameter.
Under the present normalization, the location of the zone has only a modest effect on the local contact probability, 
whereas the modelled Galactic connectivity is minimized when the civilization-density peak lies near the solar Galactocentric radius, 
where current determinations place the peak.
\end{abstract}




\section{Introduction}
\label{sec:intro}
The absence of confirmed evidence for extraterrestrial technological civilizations, despite the age and size of the Galaxy, remains one of the central problems of astrobiology \citep{Hart1975,Tarter2001}. Quantitative discussions of this problem are usually framed through the Drake equation \citep{Drake1965},
\begin{equation}
  N = DL,
  \label{eq:Drake}
\end{equation}
where $N$ is the number of communicative civilizations present in the Galaxy at a given time, $L$ is the communicative lifetime, 
and $D$ collects all remaining Drake-equation factors.
However, the existence of many civilizations does not by itself imply a high probability of interstellar contact. 
Because communication is limited in both range and duration, the problem is one of connectivity as much as abundance.

\citet{GCNT1}, hereafter Paper~I, formalized this distinction through Galactic Civilization Network Theory (GCNT), treating civilizations as nodes of a communication network and showing that the connectivity of that network is governed by a single dimensionless parameter. That analysis assumed a uniform Galactic disc. The real Galaxy is not uniform: the stellar density of the Galaxy declines approximately exponentially with Galactocentric radius, and the occurrence fraction of communicative civilizations may also vary with radius. The present paper examines how Galactic civilization networks are modified when this radial structure is restored.

\subsection{The contactability framework}\label{sec:intro:paper1}
Paper~I modelled the Galaxy as a circular disc of radius $R = 50{,}000$\,ly containing $N$ communicative civilizations distributed uniformly at random, with communication links established whenever the separation between two civilizations was smaller than $cL$, the distance a signal travels within the communicative lifetime, $c$ being the speed of light. The resulting network is a random geometric graph \citep{Penrose2003,DallChristensen2002} whose large-scale properties are governed by the single dimensionless parameter
\begin{equation}
  \eta = N \left( \frac{cL}{R} \right)^{2},
  \label{eq:eta_paper1}
\end{equation}
termed the contactability parameter.
The meaning of $\eta$ becomes explicit once the civilization abundance is expressed as a surface density.
Writing $n = N / (\pi R^{2})$ for the civilization surface density, the mean number of civilizations per unit area, Equation~\eqref{eq:eta_paper1} can be rewritten as
\begin{equation}
  \eta = \pi n (cL)^{2} ,
  \label{eq:eta_n}
\end{equation}
in which the disc radius $R$ has cancelled: $\eta$ is the expected number of civilizations lying within a distance $cL$ of a given civilization, and $\eta$ is therefore set by the local civilization density rather than by the size of the Galaxy.

Two characteristic thresholds were identified. The isolation probability follows $P_{0}\simeq e^{-\eta}$, yielding a contact threshold $\eta_{\mathrm{contact}}\simeq1$, while the appearance of a giant connected component occurs at $\eta_{\mathrm{network}}\simeq4.3$. These thresholds divide Galactic civilization networks into Isolation, Contact, and Network phases. Throughout this paper $\eta_{\mathrm{c}}$ denotes either threshold in statements that apply to both.

Combining Equation~\eqref{eq:eta_paper1} with the Drake relation gives $\eta = c^{2} D L^{3} / R^{2}$, so that contactability grows as the cube of the communicative lifetime and the contact threshold translates into a critical lifetime $L_{\mathrm{c}} = (R^{2}/c^{2}D)^{1/3}$.
Paper~I adopted the optimistic value $D \approx 0.015\,{\rm yr^{-1}}$, formed as the product of optimistic estimates of the remaining Drake factors \citep{Prantzos2020,Tsumura2020}, for which $L_{\mathrm{c}}\approx5.5\times10^{3}$\,yr and $N_{\mathrm{c}}\approx83$.

Equation~\eqref{eq:eta_n} involves only the civilization surface density, and therefore separates the two factors that determine that density.
Writing $n = \zeta \Sigma_{\star}$, where $\Sigma_{\star}$ is the stellar surface density and $\zeta$ the fraction of stars hosting communicative civilizations, gives
\begin{equation}
  \eta = \pi \zeta \Sigma_{\star} (cL)^{2}.
  \label{eq:eta_zeta}
\end{equation}
Equation~\eqref{eq:eta_zeta} reveals the principal limitation of the homogeneous model: both $\Sigma_{\star}$ and $\zeta$ are expected to vary across the Galaxy. Paper~I identified that variation as a limitation of the uniform disc, and a spatially varying contactability field, $\eta(r,\theta)$, is the natural extension of the theory. The present paper develops that extension.

\subsection{Radial structure of the stellar disc}
\label{sec:intro:structure}
The stellar surface density of the Milky Way disc is well approximated by an exponential profile with scale length $h_R \approx 2.6$\,kpc \citep{BlandHawthornGerhard2016}. Because the contactability parameter is proportional to the stellar surface density at fixed $L$ (Equation~\ref{eq:eta_zeta}), the strong radial density gradient of the Galactic disc translates directly into a radial gradient in contactability.
This variation is large compared with the contact and network thresholds, which are both quantities of order unity. A region that is marginally connected at one radius may therefore be well above the threshold further inward and well below the threshold further outward.
Connectivity in a structured Galaxy is thus not a single global state but a radial sequence of states. The natural objects of study become the radii at which these transitions occur, which we refer to as \emph{contact frontiers}.

\subsection{The Galactic habitable zone} 
\label{sec:intro:ghz} 

A second source of radial structure is the occurrence fraction $\zeta$. The idea that habitability is concentrated within a Galactic Habitable Zone (GHZ) was introduced by \citet{Gonzalez2001} and quantified by \citet{Lineweaver2004}, who combined metallicity, supernova sterilization, and evolutionary timescales to infer a preferred radius of roughly $7$--$9$\,kpc. 

The location of the GHZ remains uncertain. Some studies place the peak in the inner Galaxy \citep{Prantzos2008,Gowanlock2011}, whereas others favour the outer disc \citep{Vukotic2016,CirkovicBradbury2006}. Despite these differences, the underlying ingredients are similar: metallicity and evolutionary timescales favour the inner disc, while supernova sterilization favours the outer disc. The principal uncertainty lies in the relative weight assigned to the supernova term. 

In the present paper this uncertainty is represented by a single parameter controlling the strength of supernova sterilization. As shown in Section~\ref{sec:ghz:annulus}, varying this parameter alone shifts the peak of the civilization density from the Galactic centre to $12$\,kpc, reproducing the full range of published GHZ determinations. 

For our purposes, the GHZ is important not primarily as a habitability model but as a source of spatial structure. Because the stellar density decreases outward while the occurrence fraction is generally peaked, the resulting contactability field differs qualitatively from the contactability field of an exponential stellar disc alone. Accordingly, we examine representative inner-, mid-disc, and outer-GHZ realizations rather than adopting any single GHZ model as correct.

\subsection{Gradient percolation and the local criterion}
\label{sec:intro:gradient}
In systems with spatial density gradients, a percolation transition does not occur at a sharp boundary. Instead, the connected structure develops a finite-width frontier whose properties are governed by the theory of gradient percolation \citep{Sapoval1985,Rosso1986,Nolin2008}. Because the Galactic disc imposes exactly such a gradient on the civilization density, the boundaries between the three connectivity regimes of Paper~I cannot be expected to remain sharp. The location and width of the resulting frontiers therefore become central quantities of interest in the present work.

A second issue is the validity of the local generalization of Paper~I's criterion. At Paper~I's critical lifetime $L_{\mathrm{c}} = 5.5\times10^{3}$\,yr, the communication distance is $cL_{\mathrm{c}} = 1.7$\,kpc, corresponding to $cL/h_R \approx 0.65$, and the reference case of the present paper is barely larger. The communication range is therefore comparable to the scale over which the stellar density varies, suggesting that a purely local description based on $\eta(r)$ may not be accurate.

That concern does not preclude an exact treatment. The expected number of partners is obtained by integrating the civilization density over the communication disc, and the isolation probability follows directly from that integral. The local approximation is recovered in the limit $cL/h_R \rightarrow 0$, allowing the accuracy of the local criterion to be quantified explicitly rather than assumed.

\subsection{Objectives of this study}
\label{sec:intro:objectives}
The objective of the present study is to determine the radial structure of Galactic civilization networks. We extend the uniform-disc model of Paper~I by introducing an exponential stellar disc and a radially varying occurrence fraction based on a GHZ. The strength of the supernova sterilization term is treated as a free parameter, allowing the full range of published GHZ determinations to be represented within a single framework.

Three questions are addressed. First, where do the contact and network frontiers lie, and how wide is each frontier? Second, how are these frontiers modified by a radially varying occurrence fraction? Third, is the local generalization of Paper~I's criterion, $\eta(r)=\eta_{\rm c}$, an adequate description of a structured Galaxy, and if not, how large is the resulting error?
The connectivity thresholds of Paper~I remain unchanged. What the present analysis adds are the spatial locations of those thresholds, the widths of the corresponding transitions, and the validity limits of the local approximation.

The present paper is the second of a four-paper series on GCNT.
Paper~I develops the basic framework under the simplifying assumption that civilizations are distributed uniformly in both space and time, an idealized limit in which contactability, connectivity, and the network phase transitions appear in the simplest form.
The present paper relaxes spatial uniformity by introducing Galactic structure and GHZ models, and so determines the radial organization of civilization networks within the Milky Way.
\citet{GCNT3}, hereafter Paper~III, introduces explicit spacetime causality and carries it through to the connectivity and percolation properties of the resulting networks.
\citet{GCNT4}, hereafter Paper~IV, closes the series by relaxing the assumption common to the first three, that every civilization shares one communicative lifetime and one detection range.
The four papers together build a progressively more realistic description of Galactic civilization networks while retaining the conceptual simplicity of the original framework.

The remainder of this paper is organized as follows. Section~\ref{sec:model} introduces the structured-disc model. Section~\ref{sec:uniform} examines the radial structure produced by an exponential stellar disc alone. Section~\ref{sec:ghz} incorporates a GHZ and explores the impact of that zone on connectivity. Section~\ref{sec:discussion} discusses the implications for SETI and the Fermi paradox. Section~\ref{sec:conclusions} summarizes the results.


\section{Model} \label{sec:model}
The model of Paper~I is retained in every respect except one: the uniform disc is replaced by an axisymmetric, radially structured disc.
Civilizations are still points in a two-dimensional disc, still linked whenever the separation between two civilizations is smaller than $cL$.
What changes is the density from which the points are drawn.

\subsection{The stellar disc} \label{sec:model:disk}
The stellar surface density is taken to be exponential,
\begin{equation}
  \Sigma_{\star}(r) = \Sigma_{0}\, e^{-r/h_{R}}, \qquad r \le R_{\mathrm{tr}},
  \label{eq:expdisk}
\end{equation}
with scale length $h_R=2.6$\,kpc \citep{BlandHawthornGerhard2016}. The disc is truncated at $R_{\rm tr}=20$\,kpc, a numerical cutoff beyond which the exponential profile contributes negligibly rather than a physical edge; the consequences of that choice are quantified below and in Section~\ref{sec:disc:limits}.
To maintain consistency with Paper~I, the normalization is fixed by the local stellar surface density at the solar Galactocentric radius,
$\Sigma_{\star}(R_{0}) \simeq 80\ \mathrm{pc^{-2}}$ \citep{Bovy2017} at $R_{0} = 8.2$\,kpc, rounded from the geometric determination $R_{0} = 8.178 \pm 0.026$\,kpc \citep{GravityCollaboration2019}. This gives
\begin{equation}
  \Sigma_{0} = \Sigma_{\star}(R_{0})\, e^{R_{0}/h_{R}} \simeq 1874\ \mathrm{pc^{-2}} .
  \label{eq:sigma0}
\end{equation}
The corresponding stellar population is
\begin{equation}
  N_{\star} = \int_{0}^{R_{\mathrm{tr}}} \Sigma_{0}\, e^{-r/h_{R}}\, 2\pi r\, \mathrm{d}r
            \simeq 7.9 \times 10^{10} ,
  \label{eq:Nstar}
\end{equation}
which follows from the adopted normalization rather than constraining the normalization.
The truncation removes only $0.4\%$ of the disc and has negligible effect on the results.

\subsection{The occurrence fraction} \label{sec:model:zeta}
The fraction of stars hosting communicative civilizations is allowed to vary with radius as
\begin{equation}
  \zeta(r) = \zeta_{0}\, f_{Z}(r)\, f_{\mathrm{SN}}(r)\, f_{t}(r) ,
  \label{eq:zeta}
\end{equation}
following the factorized GHZ formalism of \citet{Lineweaver2004}. The normalization $\zeta_0$ is chosen such that $\zeta(R_0)$ remains fixed, so that all models share the same local contactability $\eta(R_0)$ and differ only in the radial shape of $\zeta(r)$.

The metallicity factor is
\begin{equation}
  f_{Z}(r) = \Phi\!\left( \frac{\mathrm{[Fe/H]}(r) + 0.5}{\sigma_{Z}} \right) ,
  \label{eq:fZ}
\end{equation}
with a Galactic abundance gradient $\mathrm{d[Fe/H]}/\mathrm{d}r=-0.06$ dex\,kpc$^{-1}$, $\sigma_Z=0.2$ dex, and a threshold $\mathrm{[Fe/H]}>-0.5$ \citep{Luck2011,Genovali2014,Lineweaver2004}. This factor favours the inner disc.

The supernova-survival factor is
\begin{equation}
  f_{\mathrm{SN}}(r) = \exp\!\left[ -k_{\mathrm{SN}}\,
                       e^{-(r-R_{0})/h_{\mathrm{SF}}} \right] ,
  \label{eq:fSN}
\end{equation}
where $h_{\mathrm{SF}}=h_R$ is adopted for simplicity. The parameter $k_{\mathrm{SN}}$ is the least constrained quantity in the model and controls the strength of supernova sterilization. As shown in Section~\ref{sec:ghz:annulus}, $k_{\mathrm{SN}}$ is also the parameter that largely determines the position of the civilization-density peak. We adopt $k_{\mathrm{SN}}=\ln2$ as a reference value and explore alternatives.

The evolutionary-timescale factor is
\begin{equation}
  f_{t}(r) = \Phi\!\left( \frac{t_{\mathrm{f}}(r) - \Delta t}{\sigma_{t}} \right) ,
  \label{eq:ft}
\end{equation}
where $t_f(r)\simeq12.5-0.45(r/\mathrm{kpc})$ Gyr, $\sigma_t=1.5$ Gyr, and $\Delta t=4$ Gyr \citep{Chiappini1997,Lineweaver2004}. This factor also favours the inner disc.

The precise functional forms are less important than the qualitative behaviour of those forms. Both $f_Z$ and $f_t$ decrease outward, whereas $f_{\mathrm{SN}}$ increases outward. As a consequence, $\zeta(r)$ typically develops a maximum at a finite radius when the outward-rising supernova-survival factor is sufficiently strong; when that factor is weak the profile instead peaks towards the Galactic centre, a case quantified in Section~\ref{sec:ghz:field}. The results that follow depend primarily on this competition between inward- and outward-favouring processes rather than on the detailed parameter values.

\subsection{The contactability field} \label{sec:model:eta}
The civilization surface density of Section~\ref{sec:intro:paper1} now varies with radius,
\begin{equation}
  n(r) = \zeta(r)\, \Sigma_{\star}(r) ,
  \label{eq:ndens}
\end{equation}
with normalization
\begin{equation}
  \int_{0}^{R_{\mathrm{tr}}} n(r)\, 2\pi r\, \mathrm{d}r = N .
  \label{eq:nnorm}
\end{equation}
The local contactability field is
\begin{equation}
  \eta(r) = \pi (cL)^{2}\, n(r) ,
  \label{eq:eta_local}
\end{equation}
which assumes that the density is approximately constant across a civilization's communication disc, and in which the communicative lifetime $L$ is common to all civilizations.

The contactability at the solar Galactocentric radius, $\eta(R_{0})$, is the free parameter throughout, and the reference case adopted below is $\eta(R_{0}) = \eta_{\mathrm{contact}} = 1$; Section~\ref{sec:uniform:eta0} sets out why.
With the disc of Section~\ref{sec:model:disk} and the fiducial Drake parameter $D = 0.015\ \mathrm{yr^{-1}}$ the reference case corresponds to $L = 6.1 \times 10^{3}$\,yr and $N = 91$.

The exact expected number of communication partners is
\begin{equation}
  \eta_{\mathrm{eff}}(r) = \int_{0}^{2\pi} \!\! \int_{0}^{cL}
     n\!\left( \sqrt{r^{2} + s^{2} + 2 r s \cos\phi} \right) s \, \mathrm{d}s \, \mathrm{d}\phi ,
  \label{eq:eta_eff}
\end{equation}
and the corresponding isolation probability is
\begin{equation}
  P_{0}(r) = \left[ 1 - \frac{\eta_{\mathrm{eff}}(r)}{N} \right]^{N-1} .
  \label{eq:P0_exact}
\end{equation}
For $N\gg1$ Equation~\eqref{eq:P0_exact} reduces to the familiar Poisson form $P_{0}\rightarrow \exp(-\eta_{\mathrm{eff}})$ \citep{MeesterRoy1996}. Equations~\eqref{eq:eta_eff} and \eqref{eq:P0_exact} therefore provide the non-local generalization of Paper~I's $P_{0}\simeq e^{-\eta}$.
Expanding $n(r)$ gives the leading correction
\begin{equation}
\begin{split}
  \frac{\eta_{\mathrm{eff}}(r)}{\eta(r)}
    &= 1 + \frac{(cL)^{2}}{8}\,\frac{\nabla^{2} n}{n}
       + \mathcal{O}\!\left[(cL)^{4}\right] \\
    &= 1 + \frac{(cL)^{2}}{8 h_{R}^{2}} \left( 1 - \frac{h_{R}}{r} \right)
       + \mathcal{O}\!\left[(cL)^{4}\right] .
\end{split}
  \label{eq:eta_expand}
\end{equation}
At $R_{0}$ the correction is only $0.1\%$ for $L=10^{3}$ yr and $4.4\%$ at the reference lifetime. The local description is therefore accurate at the few-per-cent level throughout the regime of interest.

All quantities derived from $\eta_{\mathrm{eff}}(r)$ are evaluated numerically by composite Simpson quadrature, while frontier radii are obtained by root-finding on Equations~\eqref{eq:eta_local} or \eqref{eq:eta_eff}.

\subsection{Network construction and Monte Carlo sampling} \label{sec:model:mc}
Monte Carlo realizations are required only for quantities that depend on the connected structure of the network. Civilizations are sampled from the radial density profile $2\pi r\,n(r)/N$ and assigned random azimuths. Pairs separated by less than $cL$ are connected exactly as in Paper~I, and connected components are identified using a union--find algorithm \citep{HoshenKopelman1976,NewmanZiff2001}. The giant component is defined as the largest connected component of each realization.

The principal observable is the radial giant-component fraction
\begin{equation}
  g(r) = \frac{\substack{\text{number of civilizations at radius } r \\
                          \text{belonging to the giant component}}}
              {\text{number of civilizations at radius } r} ,
  \label{eq:g_of_r}
\end{equation}
from which the frontier position $r_{50}=r(g=0.5)$ and the frontier width $w_{\mathrm{network}} = r(g\!=\!0.1) - r(g\!=\!0.9)$ are obtained by interpolation.

Because the frontier width is itself a quantity of interest, the radial bin size was tested explicitly. Bin widths of $0.05$ and $0.025$\,kpc produce statistically consistent results, whereas larger bins introduce measurable biases. All results reported below therefore use $0.05$\,kpc.

Quantities evaluated at $R_{0}$ are taken from the exact expressions of Section~\ref{sec:model:eta} rather than from binned Monte Carlo estimates. Statistical uncertainties are obtained via jackknife resampling \citep{Efron1979}, and the number of realizations is chosen such that the uncertainty in the frontier width remains below approximately $1\%$.


\section{Radial structure with a uniform occurrence fraction} \label{sec:uniform}

We first isolate the effect of the stellar disc alone, setting $\zeta(r) = \mathrm{const}$;
the occurrence fraction is restored in Section~\ref{sec:ghz}.
The model then retains two free parameters, the civilization number $N$ and the
communication distance $cL$, and throughout this section the two are tied by the Drake relation
(Equation~\eqref{eq:Drake}) at the optimistic value adopted in Paper~I,
$D = 0.015\ \mathrm{yr^{-1}}$.
One parameter is then left, and that parameter is taken to be the contactability at the solar Galactocentric radius,
$\eta(R_{0})$.

\subsection{The contactability field and the three radial zones}
\label{sec:uniform:field}

For uniform $\zeta$, the civilization density follows the stellar profile and Equation~\eqref{eq:eta_local} becomes
\begin{equation}
  \eta(r) = \pi \zeta \Sigma_{0} (cL)^{2}\, e^{-r/h_{R}} .
  \label{eq:eta_exp}
\end{equation}
The radius at which the contactability reaches a threshold value $\eta_{\rm c}$ therefore follows directly:
\begin{equation}
  r(\eta_{\mathrm{c}}) = h_{R}
    \ln\!\left( \frac{\pi \zeta \Sigma_{0} (cL)^{2}}{\eta_{\mathrm{c}}} \right) .
  \label{eq:frontier_radius}
\end{equation}
The contact and network frontiers are defined by $\eta_{\rm contact}$ and $\eta_{\rm network}$, respectively. Unlike the homogeneous model of Paper~I, the three connectivity phases are therefore represented simultaneously at different radii within the same Galaxy.

Because $\eta(r)$ decreases exponentially, the separation between the two frontiers is
\begin{equation}
  \Delta r \equiv r_{\mathrm{contact}} - r_{\mathrm{network}}
          = h_{R} \ln \eta_{\mathrm{network}} ,
  \label{eq:deltar}
\end{equation}
which is independent of $L$, $\zeta$ and $N$. Throughout this paper we adopt the asymptotic continuum-percolation value $\eta_{\rm network}=4.51$ \citep{DallChristensen2002,Penrose2003} rather than the finite-$N$ value measured in Paper~I, so that the frontier radii below depend only on the percolation threshold itself and not on the system size at which the threshold was measured; the two values differ by less than $5\%$, a difference that shifts every network radius by about $0.1$\,kpc.
Equation~\eqref{eq:deltar} then gives $\Delta r = 3.92$ kpc, or about one and a half disc scale lengths.

At the reference case $\eta(R_0)=\eta_{\rm contact}=1$ the Sun lies exactly on the contact frontier. The resulting frontier radii are
$r_{\mathrm{contact}} = R_{0} = 8.20$\,kpc and
$r_{\mathrm{network}} = R_{0} - \Delta r = 4.28$\,kpc,
which move outward by only $0.11$ and $0.06$\,kpc when the exact non-local criterion $\eta_{\rm eff}(r)=\eta_{\rm c}$ is used instead;
at $R_{0}$ the non-local correction is $\eta_{\rm eff}(R_{0})/\eta(R_{0}) = 1.044$, so that a homogeneous description of the solar neighbourhood is accurate at the few-per-cent level.

A subtle difference from Paper~I arises because a structured disc and a uniform disc contain different numbers of stars at the same local density. The resulting geometric correction is
\begin{equation}
  \frac{\eta_{\exp}(R_{0})}{\eta_{\mathrm{uniform}}}
    = \frac{R^{2} e^{-R_{0}/h_{R}}}
           {2 h_{R}^{2} \left[1-(1+R_{\mathrm{tr}}/h_{R})e^{-R_{\mathrm{tr}}/h_{R}}\right]}
    = 0.74 ,
  \label{eq:geomratio}
\end{equation}
implying only a 25\% difference in contactability, or $10\%$ in the corresponding lifetime threshold: the $L_{\mathrm{c}} = 5.5 \times 10^{3}$\,yr of Paper~I becomes $6.1 \times 10^{3}$\,yr here. How far the local criterion can be trusted once the disc is structured is taken up in Section~\ref{sec:disc:paperI}.

\subsection{The phase structure as a function of the local contactability}
\label{sec:uniform:eta0}

Nothing in Section~\ref{sec:uniform:field} requires the specific choice $\eta(R_0)=1$. Because the contactability field is exponential,
\begin{equation}
  \eta(r) = \eta(R_{0})\, e^{-(r-R_{0})/h_{R}} ,
  \label{eq:eta_of_eta0}
\end{equation}
the local value at $R_{0}$ determines the entire radial structure, including the frontier radii,
\begin{equation}
  r(\eta_{\mathrm{c}}) = R_{0} + h_{R} \ln\!\left[ \frac{\eta(R_{0})}{\eta_{\mathrm{c}}} \right] .
  \label{eq:frontier_of_eta0}
\end{equation}
Equation~\eqref{eq:frontier_of_eta0} is the natural parameterization for SETI, because $\eta(R_0)$ is the quantity directly constrained by our own non-detection.

Equation~\eqref{eq:frontier_of_eta0} immediately implies that the radial phase structure is highly sensitive to the local contactability. Each decade change in $\eta(R_0)$ shifts every frontier by
\begin{equation}
  \frac{\mathrm{d} r}{\mathrm{d} \log_{10} \eta(R_{0})} = h_{R} \ln 10 = 6.0\ \mathrm{kpc} .
  \label{eq:decade}
\end{equation}
The conditions for a Contact Zone and a Network Zone to exist anywhere in the disc are
\begin{equation}
  \eta(R_{0}) > e^{-R_{0}/h_{R}} = 0.043 ,
  \label{eq:existence_contact}
\end{equation}
and
\begin{equation}
  \eta(R_{0}) > \eta_{\mathrm{network}}\, e^{-R_{0}/h_{R}} = 0.19 .
  \label{eq:existence_network}
\end{equation}
For the fiducial Drake parameter $D=0.015\,{\rm yr^{-1}}$, these correspond to minimum communicative lifetimes of $2.1\times10^{3}$ yr and $3.5\times10^{3}$ yr, respectively.

Table~\ref{tab:eta0} summarizes the resulting phase structure. The transition is rapid: below $\eta(R_0)\simeq0.2$ no Network Zone exists anywhere in the Galaxy, while for $\eta(R_0)\lesssim0.03$ the entire disc lies in the Isolation Phase.

\begin{table*}
\centering
\caption{Radial phase structure as a function of the contactability at $R_{0}$, at
the fixed Drake parameter $D = 0.015\ \mathrm{yr^{-1}}$.
$L$ and $N$ are the communicative lifetime and civilization number that $\eta(R_{0})$
implies through the Drake relation.
$f(\eta > 1)$ and $f(\eta > 4.51)$ are the fractions of all civilizations in the disc lying
inside the contact and network frontiers; $P_{0}(R_{0})$ is the isolation probability at
$R_{0}$ and $\langle P_{0} \rangle$ is the population-weighted average of that probability over the disc, both
from Equation~\eqref{eq:P0_exact}.
A dash indicates that the zone does not exist anywhere in the disc.}
\label{tab:eta0}
\begin{tabular}{ccccccccc}
\hline
$\eta(R_{0})$ & $L$ & $N$ & $r_{\mathrm{contact}}$ & $r_{\mathrm{network}}$ &
$f(\eta\!>\!1)$ & $f(\eta\!>\!4.51)$ & $P_{0}(R_{0})$ & $\langle P_{0} \rangle$ \\
 & [yr] & & [kpc] & [kpc] & & & & \\
\hline
$2.00$ & 7649 & 115 & 10.00 & 6.09 & 0.90 & 0.68 & 0.12 & 0.09 \\
$1.00$ & 6071 &  91 & 8.20 & 4.28 & 0.83 & 0.49 & 0.35 & 0.15 \\
$0.30$ & 4064 &  61 & 5.07 & 1.15 & 0.58 & 0.07 & 0.74 & 0.36 \\
$0.10$ & 2818 &  42 & 2.21 & ---  & 0.21 & 0.00 & 0.91 & 0.63 \\
$0.03$ & 1886 &  28 & ---  & ---  & 0.00 & 0.00 & 0.97 & 0.85 \\
$0.01$ & 1308 &  20 & ---  & ---  & 0.00 & 0.00 & 0.99 & 0.95 \\
\hline
\end{tabular}
\end{table*}

Table~\ref{tab:eta0} carries one of the main observational implications of the model. In a realistic Galactic disc, the three connectivity regimes are not generic states through which every Galaxy passes. Instead, the existence of each regime depends on the local contactability. For the fiducial Drake parameter, a Galaxy possesses both Contact and Network Zones only within the narrow interval $0.2 \lesssim \eta(R_{0}) \lesssim 4.2 \times 10^{2}$, corresponding to communicative lifetimes between $3.5\times10^{3}$ and $4.6\times10^{4}$ yr.
Within the structured-disc model, a null result therefore constrains not only the local abundance of civilizations, but also the range of large-scale Galactic connectivity compatible with that local constraint.

The reference case $\eta(R_{0}) = 1$ of Section~\ref{sec:model:eta} is adopted for a different reason: $\eta(R_{0}) = 1$ is the contact threshold itself.
Setting $\eta(R_{0}) = \eta_{\mathrm{contact}}$ places the Sun exactly on the contact frontier, so that $r_{\mathrm{contact}} = R_{0}$ and every radius quoted below is measured from a frontier of known position; the choice also connects directly to Paper~I, whose critical lifetime is defined by the same condition.
The choice is a convention rather than an inference, and Equation~\eqref{eq:frontier_of_eta0} carries every result to any other value of $\eta(R_{0})$.

Our own non-detection does bound $\eta(R_{0})$, but the resulting bound is much looser than the reference case.
A model is excluded at the $5\%$ level only if that model makes our non-detection less probable than $5\%$, so the constraint is $P_{0}(R_{0}) \ge 0.05$, which is $\eta \le \ln 20 = 3.0$ in the Poisson limit and $\eta(R_{0}) \le 2.75$ with the exact form of Equation~\eqref{eq:P0_exact} along the Drake line.
The reference case, at which $P_{0}(R_{0}) = 0.35$, therefore lies a factor of $2.8$ below the observational limit rather than at that limit.
The bound is weak because our own non-detection is the outcome of a single Bernoulli trial, which is why the reference case is fixed by the threshold rather than by the data.

\subsection{The width of the contact frontier}
\label{sec:uniform:contactwidth}

A frontier radius completely characterizes the phase structure only if the underlying transition is sharp. The contact transition is not sharp.

The isolated fraction reaches a prescribed value $P_0$ at the radius where $\eta \simeq \ln(1/P_0)$. Since $\eta(r)$ decreases exponentially, the radial interval over which $P_0$ rises from $0.1$ to $0.9$ is
\begin{equation}
  w_{\mathrm{contact}}
    = h_{R} \ln\!\left[ \frac{\ln (1/0.1)}{\ln (1/0.9)} \right]
    = 3.08\, h_{R} \simeq 8.02\ \mathrm{kpc} ,
  \label{eq:w_contact}
\end{equation}
up to corrections of order $1/N$.
Equation~\eqref{eq:w_contact} contains no dependence on $L$, $N$, or $\zeta$: the width of the contact frontier is set entirely by the Galactic disc scale length. Evaluating Equation~\eqref{eq:P0_exact} at the fiducial parameters gives a width of $8.03$\,kpc, in excellent agreement with the analytic prediction.

\begin{figure}
\centering
\includegraphics[width=\columnwidth]{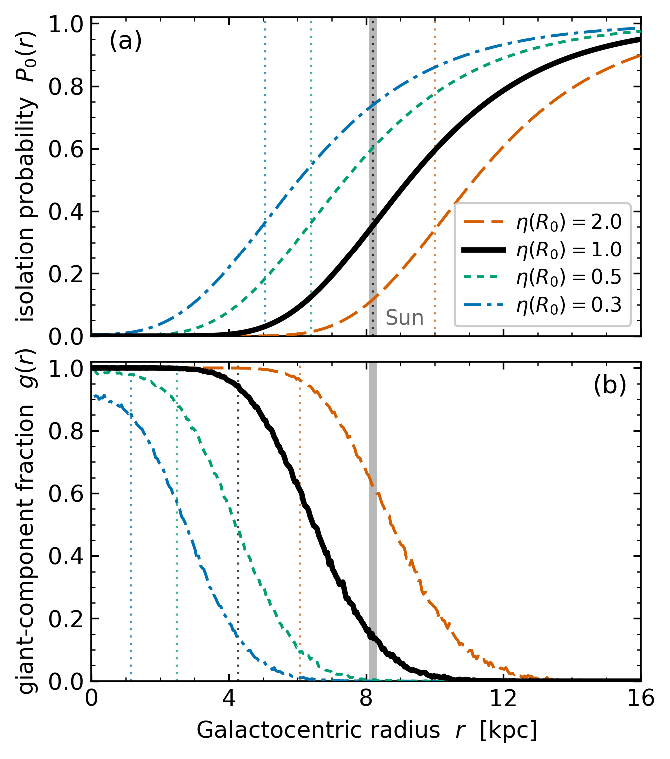}
\caption{The contact and network frontiers for several values of $\eta(R_0)$.
Panel (a) shows the isolation probability $P_0(r)$ and panel (b) the giant-component fraction $g(r)$.
The dotted vertical lines mark, for the curve of the same colour, the radius at which the local criterion of Equation~\eqref{eq:frontier_of_eta0} is met: $\eta(r) = \eta_{\mathrm{contact}} = 1$ in panel (a) and $\eta(r) = \eta_{\mathrm{network}} = 4.51$ in panel (b).
The grey vertical band marks $R_{0}$.}
\label{fig:frontiers}
\end{figure}

Panel~(a) of Figure~\ref{fig:frontiers} demonstrates this parameter independence directly: changing $\eta(R_0)$ primarily shifts the contact frontier without altering the shape of that frontier.

The broadness of the contact frontier does not arise from critical behaviour. Isolation is described by a smooth probability evaluated on a smoothly varying density field. The transition is wide simply because carrying $P_0$ from $0.1$ to $0.9$ requires $\eta$ to decrease by a factor of $\sim22$, and in an exponential disc a decrease of that size corresponds to approximately three scale lengths. The contact frontier is therefore a broad crossover extending across much of the star-forming disc rather than a sharply defined boundary.

\subsection{The width of the network frontier}
\label{sec:uniform:scaling}

Unlike the contact frontier, the network frontier is a genuine critical phenomenon whose width is governed by the percolation correlation length. In gradient percolation, balancing the correlation length $\xi \sim |\eta-\eta_{\rm c}|^{-\nu}$ against the imposed density gradient yields
\begin{equation}
  w_{\mathrm{network}} \sim cL \left( \frac{cL}{h_{R}} \right)^{-\nu/(1+\nu)}
    = (cL)^{3/7} h_{R}^{4/7} ,
  \label{eq:w_scaling}
\end{equation}
for the two-dimensional exponent $\nu=4/3$ \citep{Sapoval1985, StaufferAharony1992}.
The single independent exponent here, $\nu/(1+\nu) = 4/7$, is not phenomenological.
\citet{Nolin2008} proves for gradient percolation on the triangular lattice that the frontier is confined to a band of exactly this width --- up to an arbitrarily small correction in the exponent, and to no narrower band --- and thereby establishes rigorously the value conjectured by \citet{Sapoval1985}.
Two caveats attach to that theorem.
The proof holds on the triangular lattice, the only planar lattice for which conformal invariance of the scaling limit is known, and the theorem fixes the exponent alone.
Equation~\eqref{eq:w_scaling} therefore predicts how the width scales but not the normalization of the width.

The width must therefore be measured numerically. Using the Monte Carlo procedure of Section~\ref{sec:model:mc}, the radial profile of the giant-component fraction $g(r)$ was determined and the frontier width extracted from the radii at which $g=0.1$ and $g=0.9$. The resulting values are listed in Table~\ref{tab:width}.

\begin{table*}
\centering
\caption{The network frontier along the Drake-consistent line $D = 0.015\ \mathrm{yr^{-1}}$.
$r_{\mathrm{network}}$ is the local criterion of Equation~\eqref{eq:frontier_of_eta0};
$r_{50}$ and $w_{\mathrm{network}} = r(g\!=\!0.1) - r(g\!=\!0.9)$ are measured from the Monte Carlo profile
$g(r)$, with jackknife uncertainties; $g_{\max}$ is the largest value $g(r)$ attains at any
radius and $g_{\mathrm{tot}}$ the fraction of all civilizations in the giant component.
The radial bin width is $0.05$\,kpc.}
\label{tab:width}
\begin{tabular}{ccccccccc}
\hline
$\eta(R_{0})$ & $L$ & $N$ & $cL/h_{R}$ & $r_{\mathrm{network}}$ & $r_{50}$ & $w_{\mathrm{network}}$ &
$g_{\max}$ & $g_{\mathrm{tot}}$ \\
 & [yr] & & & [kpc] & [kpc] & [kpc] & & \\
\hline
0.3 & 4064 &  61 & 0.479 & 1.15 & $2.75 \pm 0.01$ & $4.14 \pm 0.14$ & 0.902 & 0.286 \\
0.5 & 4818 &  72 & 0.568 & 2.48 & $4.21 \pm 0.01$ & $3.71 \pm 0.02$ & 0.989 & 0.468 \\
0.7 & 5390 &  81 & 0.636 & 3.36 & $5.33 \pm 0.01$ & $3.72 \pm 0.02$ & 0.999 & 0.593 \\
1.0 & 6071 &  91 & 0.716 & 4.28 & $6.49 \pm 0.01$ & $3.89 \pm 0.03$ & 1.000 & 0.698 \\
1.5 & 6949 & 104 & 0.819 & 5.34 & $7.80 \pm 0.01$ & $4.11 \pm 0.04$ & 1.000 & 0.789 \\
2.0 & 7649 & 115 & 0.902 & 6.09 & $8.75 \pm 0.02$ & $4.24 \pm 0.03$ & 1.000 & 0.840 \\
\hline
\end{tabular}
\end{table*}

For the reference case $\eta(R_0)=1$, the measured width is $w_{\mathrm{network}} = 3.9$\,kpc, or $1.50\,h_{R}$.
Thus the Network Phase is bounded not by a sharp surface but by a band several kiloparsecs wide.

A striking result is that the width is nearly independent of position. While $\eta(R_0)$ varies by a factor of 6.7 and the frontier itself shifts by 5\,kpc, the width changes only from 3.71 to 4.24\,kpc. Table~\ref{tab:width} is not, however, a test of Equation~\eqref{eq:w_scaling}: along the Drake-consistent line $\eta(R_{0})$ varies together with $L$, so the frontier moves through different parts of the disc and the two effects cannot be separated.

A direct test is nevertheless available.
Holding $\eta(R_{0}) = 1$ fixed places the frontier at the same radius, $r_{\mathrm{network}} = 4.28$\,kpc, in every realization, while $N = 2\eta(R_{0})(h_{R}/cL)^{2}e^{R_{0}/h_{R}}$ varies as $L^{-2}$.
Scanning $cL$ from $0.70$ to $2.80$\,kpc therefore changes the civilization number by a factor of $16$, from $644$ to $40$, at fixed frontier geometry.
The measured width rises from $2.59$ to $4.67$\,kpc, a factor of $1.80$ against the $1.81$ predicted by Equation~\eqref{eq:w_scaling}, and a power-law fit gives $w_{\mathrm{network}} \propto (cL)^{0.40 \pm 0.01}$.
A width set by the size of the Network Zone would have been constant.
The frontier is therefore critical in origin rather than an artifact of the small number of civilizations that the Network Zone contains.

The fitted exponent falls about $10\%$ short of $3/7 = 0.43$, and the same scan shows why: $w_{\mathrm{network}}/h_{R}$ ranges only from $0.99$ to $1.79$, so the frontier is never thin compared with the scale over which the density varies, and a thin frontier is the limit that Equation~\eqref{eq:w_scaling} assumes.
Reaching $w_{\mathrm{network}}/h_{R} \lesssim 0.3$ would require $N \sim 2 \times 10^{5}$.
The asymptotic regime of gradient percolation is thus not accessible to a Galaxy described by the Drake relation, and the exponent is better read as an approximation than as an identity.

At very small values of $\eta(R_0)$ the transition becomes ill-defined. For example, at $\eta(R_0)=0.3$ the giant-component fraction only just reaches $0.9$ and the width is correspondingly ill-determined, indicating that only a marginal cluster is present near the Galactic centre. Although a formal Network Zone appears once $\eta(R_0)\gtrsim0.19$ (Equation~\ref{eq:existence_network}), an extended and identifiable network region requires $\eta(R_0)\gtrsim0.5$, corresponding to $L\gtrsim4.8\times10^3$ yr for the fiducial Drake parameter.

\subsection{The frontiers are intrinsically broad} 
\label{sec:uniform:broad} 
Sections~\ref{sec:uniform:contactwidth} and \ref{sec:uniform:scaling} describe transitions produced by different mechanisms --- a non-critical crossover in a decaying exponential, and a percolation transition governed by a diverging correlation length --- yet both mechanisms produce frontiers of comparable, kiloparsec-scale width.
The contact frontier spans $3.08\,h_{R}$, fixed solely by the stellar scale length, and the network frontier stays close to $1.5\,h_R$ throughout the parameter range explored here.
The broadness is a property of the exponential disc rather than a limitation of the calculation: a disc whose surface density spans three decades over the radius of that disc cannot place a threshold of order unity anywhere except across a substantial fraction of the disc. 

The Drake relation reinforces this conclusion. 
Because $\eta \propto D L^3$, maintaining a given frontier radius requires $L \propto D^{-1/3}$, and Equation~\eqref{eq:w_scaling} implies 
\begin{equation} 
w_{\mathrm{network}} \propto D^{-1/7}. 
\label{eq:w_of_D} 
\end{equation} 
Halving the width therefore requires an increase of $2^{7} = 128$ in $D$, or $2^{7.5} \simeq 180$ with the exponent measured in Section~\ref{sec:uniform:scaling}, so that even highly optimistic values of $D$ reduce the width only modestly.
The width of the network frontier is therefore set by the disc rather than by the Drake parameter, and no choice of $D$ within the models considered here makes the frontier sharp. 

The two frontiers are also sufficiently broad that the two overlap.
At the reference parameters the contact frontier runs from $6.13$ to $14.16$\,kpc and the network frontier from $4.62$ to $8.51$\,kpc, so that over $6.13$--$8.51$\,kpc --- a band containing the Sun --- both transitions are in progress at once.
This is not an accident of the parameters chosen: the nominal separation is only $\Delta r = 3.92\,{\rm kpc}$, whereas the combined half-widths already sum to $5.96$\,kpc, so the transition bands overlap throughout the range of $\eta(R_0)$ for which both frontiers exist (Figure~\ref{fig:zones}). 

\begin{figure}
\centering
\includegraphics[width=8.4cm]{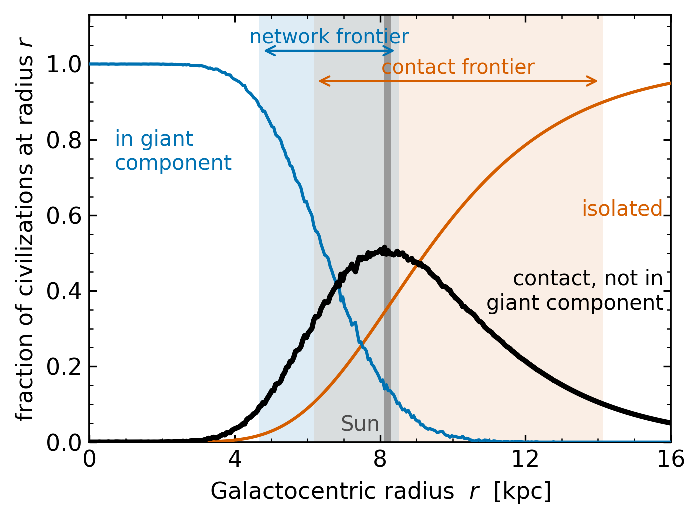} 
\caption{The three populations at the reference parameters $\eta(R_0)=1$, $L=6.1\times10^3\,{\rm yr}$ and $N=91$: the isolated fraction $P_0(r)$, the contacted but unnetworked fraction $1-P_0(r)-g(r)$, and the giant-component fraction $g(r)$. The shaded bands indicate the $10$--$90\%$ transitions of the contact and network frontiers. The overlap region contains $R_{0}$ (grey band).} 
\label{fig:phases} 
\end{figure} 

\begin{figure} 
\centering 
\includegraphics[width=8.4cm]{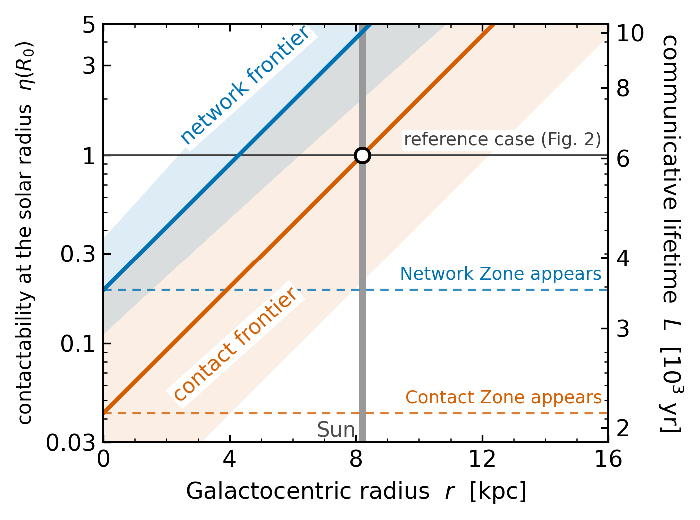} 
\caption{Radial frontiers as a function of the local contactability $\eta(R_0)$.
The solid lines are the local frontier radii of Equation~\eqref{eq:frontier_of_eta0} and the shaded bands the $10$--$90\%$ transitions of those radii, drawn in the colours of Figure~\ref{fig:phases}.
The network band is the width measured at the reference case, scaled as $w_{\mathrm{network}} \propto \eta(R_{0})^{1/7}$ and extrapolated below $\eta(R_{0}) \simeq 0.5$, where $g$ does not reach $0.9$ and the width is undefined (Section~\ref{sec:uniform:scaling}).
The horizontal dashed lines are the existence thresholds of Equations~\eqref{eq:existence_contact} and \eqref{eq:existence_network}.
The grey band marks $R_{0}$ and the open circle the reference case of Figure~\ref{fig:phases}.
The right-hand axis converts $\eta(R_{0})$ into the communicative lifetime through the Drake relation at $D = 0.015\ \mathrm{yr^{-1}}$.} 
\label{fig:zones} 
\end{figure} 

This leads to a stronger conclusion than the statement that the Galaxy is not in a single phase. 
The three phases are not sharply separated radial regions but three populations that coexist at every radius in proportions that vary continuously across the disc. 
At $R_{0}$, $35\%$ of civilizations are isolated, $50\%$ have a partner but do not belong to the giant component, and $15\%$ do belong to the giant component; that middle fraction never exceeds $0.52$ at any radius at these parameters, or $0.60$ anywhere in the range of Table~\ref{tab:width}, and nowhere are both $P_{0} < 0.1$ and $g < 0.1$ satisfied together. There is therefore no radius at which the Galaxy can be said to occupy one phase uniquely. 

The broadness of the network frontier arises because membership in the giant component is inherently non-local. 
A civilization becomes part of the Galactic network through chains of intermediate civilizations extending inward to denser regions, allowing connectivity to propagate beyond the radius where a local-density criterion would predict percolation. 
As a result, the observed frontier lies substantially outside the local criterion, and non-locality accounts for almost none of the displacement: replacing $\eta$ by $\eta_{\mathrm{eff}}$ moves the contour by $0.06$\,kpc against a displacement of $2.21$\,kpc: the displacement is a finite-gradient effect, and the local approximation becomes exact in the limit $cL/h_R \to 0$. 

An observational consequence follows immediately: at $R_{0}$ a civilization that establishes first contact need not pass through a distinct Contact Phase before reaching the Galactic network.
Because the network frontier extends well beyond the nominal location of that frontier, a civilization may move directly from the Isolation Phase into the Network Phase through contact with an already connected component. 
Whether this result survives the introduction of a radially varying occurrence fraction is examined in Section~\ref{sec:ghz}.


\section{The Galactic habitable zone} \label{sec:ghz}

Section~\ref{sec:uniform} held the occurrence fraction, $\zeta$, constant so that the
stellar disc alone would set the radial structure.
We now restore the occurrence fraction of Section~\ref{sec:model:zeta}, keeping the stellar
disc unchanged.

Equation~\eqref{eq:zeta} supplies the shape of $\zeta(r)$ but not the amplitude of $\zeta(r)$, and fixing
that amplitude $\zeta_{0}$ is the same as fixing the contactability at $R_{0}$,
since $\eta(R_{0}) = \pi \zeta(R_{0}) \Sigma_{\star}(R_{0}) (cL)^{2}$.
Section~\ref{sec:uniform} treated $\eta(R_{0})$ as the free parameter of the problem and took
$\eta(R_{0}) = 1$ as the reference case, the value at which the Sun lies on the contact
frontier, and we keep that choice here.
The communicative lifetime belonging to a given $\eta(R_{0})$ is then unchanged, and the
$L$ column of Table~\ref{tab:width} carries over intact.
What the GHZ changes is the civilization number: the Galaxy no longer holds $N = DL$ at
Paper~I's Drake parameter but a number that depends on where the zone lies, evaluated in
Section~\ref{sec:ghz:field}.
Sections~\ref{sec:uniform} and \ref{sec:ghz} therefore describe the same Galaxy with and
without a GHZ, and every difference below is attributable to the
\emph{shape} of $\zeta(r)$.

The division of labour between analysis and simulation is unchanged from
Section~\ref{sec:uniform}.
The civilization density, the position and width of the annulus, the contactability field,
the isolation probability and every zone edge quoted below follow in closed form or by
quadrature; the Monte Carlo construction of Section~\ref{sec:model:mc} is needed again only
where the giant component is involved, namely in Section~\ref{sec:ghz:network} and in the last
column of Table~\ref{tab:ghz}.

\subsection{The civilization density becomes an annulus} 
\label{sec:ghz:annulus} 

The stellar surface density decreases monotonically with radius, whereas the occurrence fraction $\zeta(r)$ is constructed from factors that act in opposite directions. Metallicity and evolutionary time favour the inner disc, while supernova survival favours the outer disc. The civilization surface density $n(r)=\zeta(r)\Sigma_{\star}(r)$ is therefore peaked at a finite radius whenever supernova sterilization is strong enough to overcome the inward-rising factors. 

Since only the supernova term varies exponentially, while the metallicity and evolutionary-time terms vary comparatively slowly, the condition ${\rm d}\ln n/{\rm d}r=0$ yields $r_{\rm peak}=R_0+h_{\rm SF}\ln(k_{\rm SN}h_R/h_{\rm SF})$, which for the choice $h_{\rm SF}=h_R$ adopted in Section~\ref{sec:model:zeta} becomes
\begin{equation} 
r_{\rm peak}=R_0+h_R\ln k_{\rm SN} . 
\label{eq:rpeak} 
\end{equation} 

With the metallicity gradient, the evolutionary-time factor and $h_{\rm SF}$ held at the values of Section~\ref{sec:model:zeta}, the peak position is therefore governed by $k_{\rm SN}$ alone, and only logarithmically. Table~\ref{tab:ksn} shows that Equation~\eqref{eq:rpeak} reproduces the exact profile to better than $0.04$\,kpc for $k_{\rm SN}\lesssim1$, the agreement degrading only where the metallicity and time factors cease to be slowly varying at the peak. The annular structure is therefore a generic consequence of the competing radial trends in habitability rather than a tuned feature of the model. Equation~\eqref{eq:rpeak} also provides a simple interpretation of the GHZ literature. Varying $k_{\rm SN}$ alone, and holding every other factor fixed, moves the peak across the whole published range: small values reproduce inner-Galaxy GHZ determinations \citep{Prantzos2008,Gowanlock2011}, intermediate values reproduce the classical 7--9 kpc annulus of \citet{Lineweaver2004}, and large values reproduce the outer-disc peaks found by \citet{Vukotic2016}. The diversity of published GHZ locations can therefore be understood largely as different assumptions regarding the strength of supernova sterilization. 

\begin{table*}
\centering
\caption{Position of the peak of the civilization density as a function of the supernova
coefficient $k_{\mathrm{SN}}$ of Equation~\eqref{eq:fSN}, with all other parameters as in
Section~\ref{sec:model:zeta}.
$f_{\mathrm{SN}}(R_{0}) = e^{-k_{\mathrm{SN}}}$ is the corresponding survival probability at $R_{0}$.
The peak is obtained from the exact profile; Equation~\eqref{eq:rpeak} reproduces the entries
to $0.04$\,kpc for $k_{\mathrm{SN}} \le 1$, and to $0.13$ and $0.41$\,kpc at
$k_{\mathrm{SN}} = 2$ and $5$.}
\label{tab:ksn}
\begin{tabular}{lcccccccc}
\hline
$k_{\mathrm{SN}}$ & 0.05 & 0.10 & 0.20 & 0.40 & $\ln 2$ & 1.00 & 2.00 & 5.00 \\
$f_{\mathrm{SN}}(R_{0})$ & 0.95 & 0.90 & 0.82 & 0.67 & 0.50 & 0.37 & 0.14 & 0.01 \\
$r_{\mathrm{peak}}$ [kpc] & 0.41 & 2.21 & 4.01 & 5.81 & 7.23 & 8.16 & 9.87 & 11.98 \\
\hline
\end{tabular}
\end{table*}

\begin{figure*}
\centering
\includegraphics[width=\textwidth]{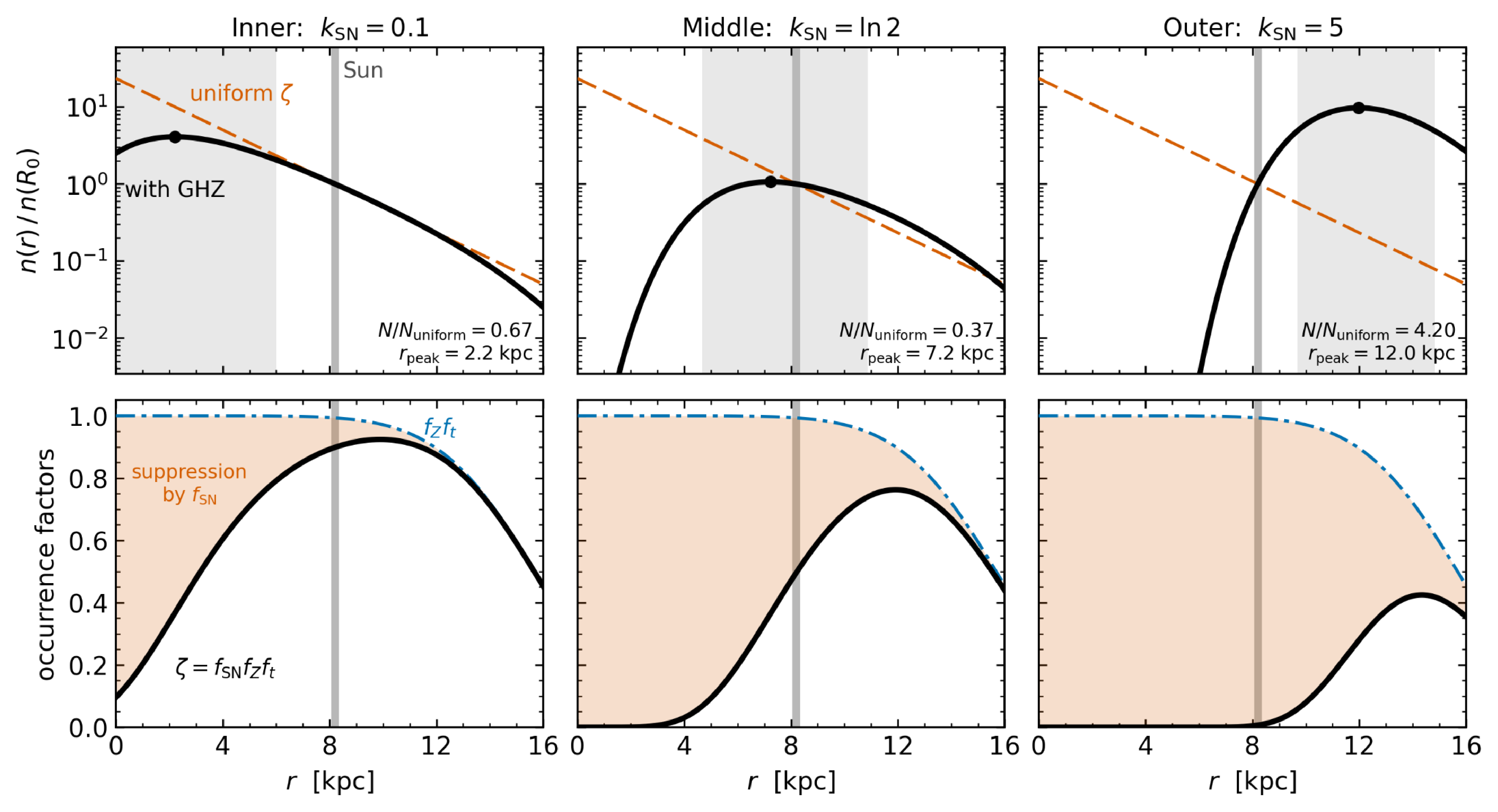}
\caption{The civilization density and the ingredients that produce that density, for the Inner, Middle
and Outer zones of Section~\ref{sec:ghz:annulus}, with $k_{\mathrm{SN}} = 0.1$, $\ln 2$ and
$5$.
Only $f_{\mathrm{SN}}$ differs between the columns.
Upper row: the civilization surface density $n(r) = \zeta(r)\Sigma_{\star}(r)$ (heavy black,
with the peak marked) and the uniform-$\zeta$ profile (orange dashed), both at the same local
value $n(R_{0})$, on a logarithmic scale; the light band is the full width at half maximum of
$n(r)$ in that column.
Lower row: the occurrence fraction $\zeta = f_{\mathrm{SN}} f_{Z} f_{t}$ (black) and the
combined metallicity and time factors $f_{Z} f_{t}$ of Equations~\eqref{eq:fZ} and
\eqref{eq:ft} (blue dash-dotted), on a linear scale, the orange shaded area between the two curves
being the suppression by the supernova factor of Equation~\eqref{eq:fSN}.
The grey vertical band marks $R_{0}$.
$N/N_{\mathrm{uniform}}$ is the total civilization number relative to the
uniform-$\zeta$ disc at the same $n(R_{0})$.}
\label{fig:ghzprofile}
\end{figure*}

Figure~\ref{fig:ghzprofile} shows three representative cases, which we refer to as the \emph{Inner}, \emph{Middle}, and \emph{Outer} zones, corresponding to $k_{\rm SN}=0.1$, $\ln2$, and $5$. We adopt the Middle zone as the reference case throughout the remainder of this paper, while carrying the Inner and Outer zones in parallel to span the full range of published GHZ determinations. 

A GHZ also changes the total number of civilizations, since the star-weighted mean occurrence fraction differs between the three zones. This dependence is not monotonic: strong suppression of the inner disc can be overcompensated by an enhanced outer population, so that some GHZ realizations contain more civilizations than the corresponding uniform-$\zeta$ model despite sharing the same local occurrence fraction. 

Sections~\ref{sec:ghz:field}--\ref{sec:ghz:network} carry the three zones through in parallel, Table~\ref{tab:ghz} summarizes the resulting quantities, and Figure~\ref{fig:ghzphases} compares the network structures of the three zones directly.

\subsection{The contactability field flattens}
\label{sec:ghz:field}

Because $\eta(r) = \pi (cL)^{2} n(r)$, the contactability field inherits the shape of the
civilization density, and because the occurrence fraction is normalized at $R_{0}$
the natural way to read the field is relative to the solar radius,
\begin{equation}
  \frac{\eta(r)}{\eta(R_{0})}
    = \frac{\zeta(r)}{\zeta(R_{0})}\, e^{-(r-R_{0})/h_{R}} .
  \label{eq:etashape}
\end{equation}
Because the normalization holds $\zeta(R_{0})$ fixed, all four Galaxies considered here ---
the uniform-$\zeta$ disc of Section~\ref{sec:uniform} and the three zones --- share the same
$\eta(R_{0})$ and the same communicative lifetime, and differ only in the radial shape of
$\zeta(r)$ and hence in the total civilization number.
Table~\ref{tab:ref} collects the reference-case values on which this section and the two that
follow are based.
At $\eta(R_{0}) = 1$ the radius $R_{0}$ is by construction an edge of the Contact Zone in every
model, so the four Galaxies differ only in where the remainder of that zone lies.

\begin{table*}
\centering
\caption{The reference case, $\eta(R_{0}) = \eta_{\mathrm{contact}} = 1$, for the
uniform-$\zeta$ disc of Section~\ref{sec:uniform} and the three GHZ zones of
Section~\ref{sec:ghz:annulus}.
All four Galaxies share the same $\eta(R_{0})$, the same $\zeta(R_{0})$ and the same
communicative lifetime $L = 6.1 \times 10^{3}$\,yr.
$\eta_{\max}$ is the maximum of $\eta(r)$ over the disc and $\eta_{\mathrm{eff}}(R_{0})$ the
exact non-local value at $R_{0}$ from Equation~\eqref{eq:eta_eff}.
Both are listed as ratios to $\eta(R_{0})$ because, alone among the entries, these two ratios are
unchanged if a different $\eta(R_{0})$ is adopted; at the reference case the ratios equal the
quantities themselves.
$g(R_{0})$ and $g_{\mathrm{tot}}$ are the giant-component fraction at $R_{0}$ and
over the whole disc; the last row is the probability that a contacted civilization at
$R_{0}$ already belongs to the giant component.
The zones are the radial intervals in which $\eta(r)$ exceeds $\eta_{\mathrm{contact}} = 1$
and $\eta_{\mathrm{network}} = 4.51$.
A dash indicates that the quantity does not apply or that the zone does not exist.}
\label{tab:ref}
\begin{tabular}{lcccc}
\hline
 & uniform $\zeta$ & \textsc{Inner} & \textsc{Middle} & \textsc{Outer} \\
\hline
$k_{\mathrm{SN}}$                        & ---     & $0.1$   & $\ln 2$ & $5$     \\
$r_{\mathrm{peak}}$ [kpc]                & ---     & $2.2$   & $7.2$   & $12.0$  \\
$\langle \zeta \rangle / \zeta(R_{0})$   & $1$     & $0.667$ & $0.370$ & $4.199$ \\
$N$                                      & $91$    & $61$    & $34$    & $382$   \\
$\eta_{\max}/\eta(R_{0})$                & $23.4$  & $4.09$  & $1.07$  & $9.73$  \\
$\eta_{\mathrm{eff}}(R_{0})/\eta(R_{0})$ & $1.044$ & $1.026$ & $0.954$ & $1.685$ \\
Contact Zone [kpc]  & $0$--$8.20$ & $0$--$8.20$ & $6.36$--$8.20$ & $8.20$--$17.40$ \\
Network Zone [kpc]  & $0$--$4.28$ & ---         & ---            & $9.57$--$15.00$ \\
$P_{0}(R_{0})$                           & $0.354$ & $0.360$ & $0.387$ & $0.186$ \\
$g(R_{0})$                               & $0.151$ & $0.073$ & $0.191$ & $0.647$ \\
$g_{\mathrm{tot}}$                       & $0.698$ & $0.349$ & $0.122$ & $0.842$ \\
$g(R_{0})/[1-P_{0}(R_{0})]$              & $0.23$  & $0.11$  & $0.31$  & $0.80$  \\
\hline
\end{tabular}
\end{table*}

Three properties of Equation~\eqref{eq:etashape} organize the rest of the section, and only the
first property is common to all three zones of Section~\ref{sec:ghz:annulus}.

The first is the flattening.
Over the annulus that contains the central $90\%$ of the civilizations, the bare stellar disc
spans a factor of $74$ in $\eta$; with a GHZ the same measure gives $20$, $8.7$ and
$6.0$ for $k_{\mathrm{SN}} = 0.1$, $\ln 2$ and $5$.
The occurrence fraction rises outward over precisely the range in which the stellar disc falls,
so the zone flattens the field wherever the civilizations actually are, and does so whatever the
position of the peak.
That flattening is what makes the two thresholds decisive: both thresholds are numbers of order
unity, and there is no longer a region of the disc that clears the thresholds by orders of
magnitude while the solar radius does not.

The second is that the phases are not always annuli.
Since $\zeta$ is dominated by $f_{\mathrm{SN}}$ near the centre, Equation~\eqref{eq:etashape}
gives $\eta(0)/\eta(R_{0}) = e^{R_{0}/h_{R}}
\exp[-k_{\mathrm{SN}}(e^{R_{0}/h_{\mathrm{SF}}} - 1)]$, so the Galactic centre remains above
the contact threshold whenever
\begin{equation}
  k_{\mathrm{SN}} < \frac{\ln\!\left[ \eta(R_{0})\, e^{R_{0}/h_{R}} \right]}
                         {e^{R_{0}/h_{\mathrm{SF}}} - 1} ,
  \label{eq:kdisk}
\end{equation}
which at $\eta(R_{0}) = 1$ is $k_{\mathrm{SN}} < 0.141$.
For the inner-Galaxy determination the Contact Zone therefore still reaches the centre and is a
disc bounded only on the outside, leaving the radial ordering of Section~\ref{sec:uniform:field}
unchanged.
Only above that value does the Contact Zone acquire an inner edge, so that the Isolation Phase
occupies the Galactic centre as well as the outer disc and the ordering is inverted.

The third property governs everything below: how much more contactable the best radius is
than the solar radius.
The maximum of Equation~\eqref{eq:etashape} is $4.09$ at $2.21$\,kpc for
$k_{\mathrm{SN}} = 0.1$, $1.066$ at $7.23$\,kpc for $\ln 2$, and $9.73$ at $11.98$\,kpc for
$5$.
The ratio $\eta_{\max}/\eta(R_{0})$ has an exact minimum: with $h_{\mathrm{SF}} = h_{R}$ the peak
of the civilization density sits at $R_{0}$ when $k_{\mathrm{SN}} = 1$, the ratio is then exactly
unity, and the ratio rises on both sides, the further in either direction the zone is placed
from $R_{0}$ the larger the ratio becomes.
The existence conditions follow directly, since a Contact Zone requires
$\eta(R_{0}) > [\eta_{\max}/\eta(R_{0})]^{-1}$ and a Network Zone
$\eta(R_{0}) > 4.51\,[\eta_{\max}/\eta(R_{0})]^{-1}$; both are listed in Table~\ref{tab:ghz}.
Sections~\ref{sec:ghz:solar} and \ref{sec:ghz:network} work out what the minimum costs.

\begin{table*}
\centering
\caption{Radial phase structure with a GHZ, for the Inner, Middle and
Outer zones of Section~\ref{sec:ghz:annulus} and at the same communicative lifetimes as
Table~\ref{tab:width}, which gives the uniform-$\zeta$ Galaxy at the same $\eta(R_{0})$.
The local occurrence fraction is held fixed in all three, so $N$ differs from
Table~\ref{tab:width} by the factor $\langle \zeta \rangle / \zeta(R_{0})$ given in each block
heading.
$\eta_{\max}$ is the maximum of the contactability field over the disc; that maximum and the zone edges
are obtained from Equation~\eqref{eq:etashape} with $\eta(r)$ normalized exactly to the
$\eta(R_{0})$ of the first column, whereas the integer $N$ listed enters only $P_{0}(R_{0})$
and the Monte Carlo.
The zones are the radial intervals within which $\eta(r)$ exceeds
$\eta_{\mathrm{contact}} = 1$ and $\eta_{\mathrm{network}} = 4.51$, a lower edge of $0$
meaning that the zone reaches the Galactic centre and is a disc rather than an annulus.
$P_{0}(R_{0})$ is from Equation~\eqref{eq:P0_exact}; $g_{\mathrm{tot}}$ is the fraction of all
civilizations in the giant component, from the Monte Carlo, with jackknife uncertainties,
which are not quoted where the uncertainties fall below $0.001$.
A dash indicates that the phase does not occur anywhere in the disc.}
\label{tab:ghz}
\begin{tabular}{cccccccc}
\hline
$\eta(R_{0})$ & $L$ & $N$ & $\eta_{\max}$ & Contact Zone & Network Zone &
$P_{0}(R_{0})$ & $g_{\mathrm{tot}}$ \\
 & [yr] & & & [kpc] & [kpc] & & \\
\hline
\multicolumn{8}{l}{\textsc{Inner} zone: $k_{\mathrm{SN}} = 0.1$, $r_{\mathrm{peak}} = 2.2$\,kpc, $\langle \zeta \rangle / \zeta(R_{0}) = 0.667$} \\
0.5 &  4818 &  48 &  2.05 & $0$--$6.09$    & ---              & 0.607 & $0.159 \pm 0.005$ \\
1.0 &  6071 &  61 &  4.09 & $0$--$8.20$    & ---              & 0.360 & $0.349 \pm 0.006$ \\
2.0 &  7649 &  76 &  8.19 & $0$--$10.07$   & $0$--$5.68$      & 0.126 & $0.719 \pm 0.007$ \\
4.0 &  9637 &  96 & 16.38 & $0$--$11.77$   & $0$--$7.86$      & 0.014 & $0.897 \pm 0.002$ \\
8.0 & 12141 & 121 & 32.75 & $0$--$13.26$   & $0$--$9.76$      & $<0.001$ & $0.963 \pm 0.002$ \\
\hline
\multicolumn{8}{l}{\textsc{Middle} zone: $k_{\mathrm{SN}} = \ln 2$, $r_{\mathrm{peak}} = 7.2$\,kpc, $\langle \zeta \rangle / \zeta(R_{0}) = 0.370$} \\
0.5 &  4818 &  27 &  0.53 & ---            & ---              & 0.621 & $0.100 \pm 0.003$ \\
1.0 &  6071 &  34 &  1.07 & $6.36$--$8.20$ & ---              & 0.387 & $0.122 \pm 0.002$ \\
2.0 &  7649 &  42 &  2.13 & $4.59$--$11.06$& ---              & 0.159 & $0.178 \pm 0.008$ \\
4.0 &  9637 &  53 &  4.26 & $3.79$--$12.85$& ---              & 0.028 & $0.328 \pm 0.016$ \\
8.0 & 12141 &  67 &  8.53 & $3.23$--$14.28$& $4.77$--$10.70$  & 0.001 & $0.768 \pm 0.022$ \\
\hline
\multicolumn{8}{l}{\textsc{Outer} zone: $k_{\mathrm{SN}} = 5$, $r_{\mathrm{peak}} = 12.0$\,kpc, $\langle \zeta \rangle / \zeta(R_{0}) = 4.199$} \\
0.5 &  4818 & 304 &  4.87 & $8.71$--$16.44$& $11.15$--$12.87$ & 0.484 & $0.202 \pm 0.006$ \\
1.0 &  6071 & 382 &  9.73 & $8.20$--$17.40$& $9.57$--$15.00$  & 0.186 & $0.842 \pm 0.006$ \\
2.0 &  7649 & 482 & 19.47 & $7.79$--$18.23$& $8.82$--$16.25$  & 0.018 & $0.985 \pm 0.001$ \\
4.0 &  9637 & 607 & 38.93 & $7.45$--$18.97$& $8.28$--$17.25$  & $<0.001$ & $0.997$ \\
8.0 & 12141 & 765 & 77.86 & $7.15$--$19.64$& $7.86$--$18.10$  & $<0.001$ & $1.000$ \\
\hline
\end{tabular}
\end{table*}

\subsection{Our own prospects move least} 
\label{sec:ghz:solar} 
The local contactability at $R_{0}$ is fixed by construction, so the introduction of a GHZ has only a modest effect on our own prospects for contact. 
At the reference case, the isolation probability at $R_{0}$ is $P_{0}(R_{0})=0.360$ (Inner), $0.387$ (Middle), and $0.186$ (Outer), compared with $0.354$ for the corresponding uniform-$\zeta$ Galaxy. 
Thus, even relocating the peak of the civilization density by almost 10\,kpc changes the local probability of isolation by less than a factor of two. 

\begin{figure*}
\centering
\includegraphics[width=\textwidth]{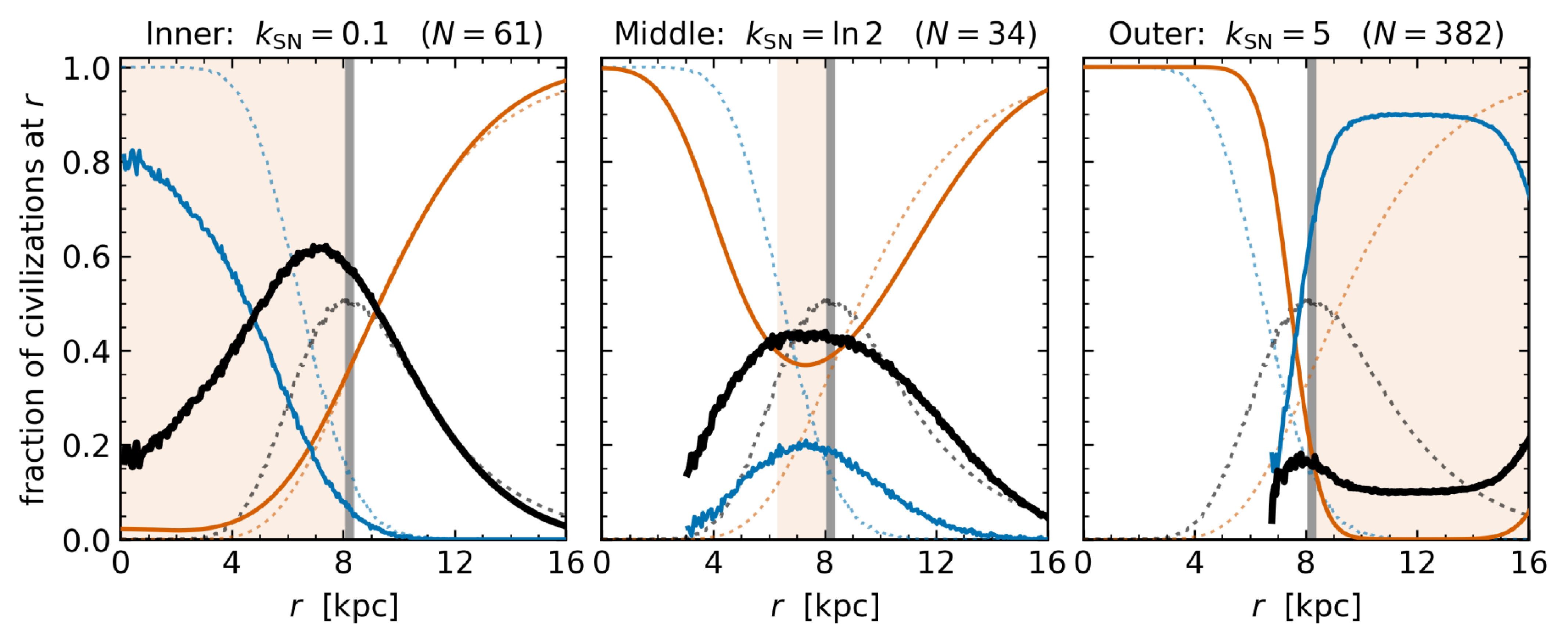}
\caption{The three populations at the reference case $\eta(R_{0}) = 1$,
$L = 6.1 \times 10^{3}$\,yr, for the Inner, Middle and Outer zones of
Section~\ref{sec:ghz:annulus}.
The colour coding is that of Figure~\ref{fig:phases}: orange is the isolated fraction
$P_{0}(r)$, heavy black the contacted but unnetworked fraction $1 - P_{0}(r) - g(r)$, and blue
the giant-component fraction $g(r)$.
Solid curves are the GHZ Galaxy at the civilization number $N$ given in each
panel; dotted curves repeat the uniform-$\zeta$ case of Figure~\ref{fig:phases} at the same
$\eta(R_{0})$ and the same $L$, for which $N = 91$.
The shaded band is the Contact Zone, the interval within which $\eta(r) > 1$, and the grey
vertical band marks $R_{0}$.}
\label{fig:ghzphases}
\end{figure*}

The radial structure of the isolated fraction, however, changes qualitatively. Under a uniform occurrence fraction, $P_0(r)$ decreases monotonically towards the Galactic centre. With a GHZ, the isolated fraction instead develops a minimum near the density peak and rises in both directions. Figure~\ref{fig:ghzphases} shows the resulting population structure for the three representative zones. 

Whether this weak local sensitivity resolves the Fermi paradox depends strongly on where the GHZ is located. In the uniform model, the inner Galaxy is much more strongly connected than the solar neighbourhood, so our continued non-detection can be interpreted as a local circumstance. The Middle zone largely removes this escape, because the most contactable region is more contactable than our own by only $7\%$. 
In contrast, the Inner and Outer zones preserve regions that are substantially more connected, by factors of $4.1$ and $9.7$, respectively. The bound $\eta(R_{0}) \lesssim 3$ of Section~\ref{sec:uniform:eta0} therefore becomes a statement about the entire Galaxy only when the GHZ is centred close to $R_{0}$. 

The GHZ also separates two distinct questions: where civilizations are most numerous and where individual stars are most likely to host civilizations. Because the civilization density follows $n=\zeta\Sigma_\star$, the peak of the civilization population lies at $r_{\rm peak}$, whereas the occurrence fraction itself peaks farther outward. For observational SETI the position of the civilization-density peak is generally more relevant, since that position determines where potential communication partners are most abundant. Which direction is the better target follows from the position of the peak alone.
A survey directed towards the Galactic centre collects more civilizations than one of equal depth directed towards the anticentre whenever the civilization density is still rising inward at $R_{0}$, which by Equation~\eqref{eq:rpeak} means $r_{\rm peak} < R_{0}$, that is $k_{\rm SN} < 1$.
The uniform-$\zeta$ disc and the Inner zone therefore favour the centre, the Outer zone favours the anticentre, and the Middle zone, whose peak lies about a kiloparsec inside $R_{0}$, gives little preference either way.
Increasing the survey depth moves the crossing somewhat below $k_{\rm SN} = 1$, since an inward beam that passes $r_{\rm peak}$ then enters the sterilized inner disc, but the ordering does not change. A GHZ therefore does not tell us to look outward so much as a GHZ removes the reason to look inward, and only the Outer zone makes the anticentre more favourable than the Galactic centre within the model; a stronger form of that conclusion has been argued on different grounds by \citet{CirkovicBradbury2006}. Nevertheless, detectability remains dominated by distance, so nearby targets continue to be favoured \citep{TurnbullTarter2003a,TurnbullTarter2003b}. The primary effect of the GHZ is therefore not to redirect SETI searches, but to alter how null detections should be interpreted in terms of Galactic connectivity.

\subsection{Whether the network survives depends on where the zone lies} 
\label{sec:ghz:network} 

The giant component is far more sensitive to the position of the GHZ than the isolation probability. At the reference case, the fraction of civilizations belonging to the largest connected component is $g_{\mathrm{tot}} = 0.349$ (Inner), $0.122$ (Middle), and $0.842$ (Outer), compared with $0.698$ for the corresponding uniform-$\zeta$ Galaxy (Table~\ref{tab:ghz}). Whereas the isolation probability at $R_{0}$ changes by less than a factor of two across these models, the networked fraction of the Galaxy changes by a factor of seven.

The reason is that contact is fundamentally local, whereas network formation is not local. The local occurrence fraction at $R_{0}$ is fixed by construction, so the contact threshold is only weakly affected by the GHZ. A giant component, however, requires that some region of the disc exceed $\eta_{\mathrm{network}}$. The Inner zone preserves a highly connected inner Galaxy, while the Outer zone creates a new highly connected annulus at large radius. Only the Middle zone fails to produce any region above the network threshold for $\eta(R_0)=1$. As a result, a Network Zone appears only when $\eta(R_0)\gtrsim4.2$, approximately twenty times the value required in the corresponding uniform disc and at the very edge of the range that our own non-detection permits.

This behaviour follows from the minimum of $\eta_{\max}/\eta(R_0)$ identified in Section~\ref{sec:ghz:field}. Because that ratio stays below $\eta_{\mathrm{network}}$ over the whole band $0.090 < k_{\mathrm{SN}} < 3.96$, no Network Zone exists anywhere in the disc at $\eta(R_0) = 1$ for any zone peaking between $2$ and $11$\,kpc. A GHZ centred near $R_{0}$ caps the contactability of the entire Galaxy at the value measured locally, so that our own non-detection bounds the Galaxy rather than the solar neighbourhood. The Middle zone is therefore the most restrictive configuration, and is also the configuration closest to the GHZ determination of \citet{Lineweaver2004}.

These comparisons are statements about the reference contactability, and the scope should be stated plainly. The ratio $\eta_{\max}/\eta(R_{0})$ is itself independent of $\eta(R_{0})$, but whether the resulting maximum clears $\eta_{\mathrm{network}}$ is not. Table~\ref{tab:ghz} shows that the Middle zone does acquire a Network Zone, running from $4.77$ to $10.70$\,kpc with $g_{\mathrm{tot}} = 0.77$, once $\eta(R_{0})$ reaches $8$. What is peculiar to the Middle zone is therefore not the absence of a giant component at every density, but the persistence of that absence across the whole range of $\eta(R_{0})$ that our own non-detection leaves open, $\eta(R_{0}) \lesssim 3$ by Section~\ref{sec:uniform:eta0}.

An observational consequence concerns first contact. The fraction of contacted civilizations at $R_{0}$ that already belong to the giant component is $0.11$ (Inner), $0.31$ (Middle), and $0.80$ (Outer), compared with $0.23$ in the uniform-$\zeta$ model. Thus the probability that first contact provides access to a Galaxy-scale network depends strongly on the position of the GHZ, even though the probability of contact itself changes only modestly. 

The overall picture is therefore simple. A GHZ leaves our own prospects for contact close to those predicted by the homogeneous model, because those prospects depend primarily on the local occurrence fraction. What the GHZ controls is the connectivity of the rest of the Galaxy. A zone centred near $R_{0}$ does not greatly reduce our chance of contact; rather, such a zone removes the existence of a large Galactic network into which that contact might lead.


\section{Discussion} \label{sec:discussion}

\subsection{A Galaxy without a boundary} \label{sec:disc:boundary} Sections~\ref{sec:uniform:broad} and \ref{sec:ghz:field} established the widths of the frontiers numerically; the present section sets out what those widths mean. The uniform-$\zeta$ Galaxy of Section~\ref{sec:uniform} and the GHZ Galaxies of Section~\ref{sec:ghz} are governed by different mechanisms, yet both classes of model produce the same structure. In the uniform case, the exponential stellar disc is too shallow to localize the connectivity transitions sharply, while in the GHZ models the occurrence fraction partially cancels the stellar-density gradient. Neither the contact frontier nor the network frontier forms a sharp boundary in either case. The consequence is conceptual rather than quantitative: the question of whether humanity lies inside or outside an inhabited region presupposes a boundary that these models do not contain, and the question therefore has no well-defined answer within them. That conclusion requires only an exponential stellar disc and connectivity thresholds of order unity, and is largely independent of the additional assumptions discussed in Sections~\ref{sec:disc:arms} and~\ref{sec:disc:limits}. Whether a Galaxy built on qualitatively different density models could sustain a sharp boundary is not addressed here.

\subsection{The status of Paper~I} \label{sec:disc:paperI} A central question posed in Section~\ref{sec:intro} was whether the local generalization of Paper~I's criterion provides an adequate description of a structured Galaxy. The answer is mixed. The connectivity thresholds themselves are unchanged, because the thresholds are properties of random geometric graphs rather than properties of the Milky Way. What changes is the radial location and width of the corresponding frontiers. For quantities that depend only on local contactability, the homogeneous approximation remains accurate. The correction factor $\eta_{\mathrm{eff}}(R_0)/\eta(R_0)$ differs from unity by less than $5\%$ for the uniform-$\zeta$ disc and for the Inner and Middle GHZ models, the Outer zone being the exception: there the Sun sits on a steeply rising flank, the factor is $1.685$, and a homogeneous description of the solar neighbourhood is wrong by more than two thirds. Accordingly, the contact threshold derived in Paper~I remains a good description of the solar neighbourhood. The situation is different for network connectivity. Membership in the giant component is intrinsically non-local. Galaxies that are identical at $R_{0}$ by construction can differ by a factor of seven in the fraction of civilizations belonging to the Galactic network. Consequently, the network threshold cannot be interpreted purely locally. Paper~I should therefore be viewed as a local description of contactability at $R_{0}$, whereas the present work provides the radial structure that this local description averages over.

\subsection{The Fermi paradox}
\label{sec:disc:fermi}

Our own non-detection constrains the local contactability $\eta(R_0)$, and the traditional escape from the Fermi paradox is to argue that the local value need not reflect the rest of the Galaxy: the Galaxy may be well connected overall while the solar neighbourhood happens to be unusually isolated.

The uniform-disc model of Section~\ref{sec:uniform} makes that escape plausible. At $\eta(R_0)=1$, the contactability of the Galactic centre exceeds the contactability at $R_{0}$ by a factor of $23.4$, so the inner Galaxy can remain strongly connected even when $R_{0}$ lies near the limit of the Contact Phase.

The GHZ models substantially reduce this freedom. The maximum contactability relative to the solar neighbourhood is reduced to factors of $4.1$, $1.07$, and $9.7$ for the Inner, Middle, and Outer zones, respectively. Only the Middle zone effectively removes the escape, because the most contactable region is barely more contactable than the solar neighbourhood. In that case, a non-detection at $R_{0}$ becomes closer to a Galaxy-wide constraint.

The Fermi paradox is therefore not removed by Galactic structure, but the paradox does not become entirely local either. The quantity constrained by observation, $\eta(R_0)$, is also the quantity governing our own prospects for contact, while the remaining uncertainty is encoded in the radial structure of $\zeta(r)$. A Galaxy can be more connected than the solar neighbourhood only by the limited factor permitted by the GHZ and only at the radius where the GHZ places the peak.

\subsection{Spiral arms}
\label{sec:disc:arms}

The models considered in Sections~\ref{sec:uniform} and~\ref{sec:ghz} are axisymmetric, whereas the Milky Way contains spiral structure. To assess the importance of spiral structure, we introduced a spiral perturbation of the form
\begin{equation}
  n(r,\theta) = \bar{n}(r)\,\bigl[ 1 + A\,F(\psi) \bigr] , \qquad
  \psi = m \left( \theta - \frac{1}{\tan p} \ln \frac{r}{R_{0}} \right) ,
  \label{eq:arm}
\end{equation}
where the azimuthal mean remains unchanged so that the radial profile, civilization number, and $\eta(R_0)$ are identical to the axisymmetric case.

For realistic Milky Way arm--interarm contrasts in the old stellar population, the resulting giant-component fractions are indistinguishable from the axisymmetric model within the Monte Carlo uncertainties.
This holds both for the near-infrared surface-brightness ratio of $1.32$ measured by \citet{DrimmelSpergel2001}, which is the appropriate comparison for a surface-density model, and for the larger mid-plane density ratio of $2.3$ reported by the same authors. The same conclusion is consistent with arm contrasts observed in external spiral galaxies \citep{RixZaritsky1995}.

Spiral structure begins to affect connectivity only for contrasts $\gtrsim 5$, substantially larger than those observed in the Milky Way. In that regime the giant component is reduced rather than enhanced, because civilizations become concentrated into narrow ribbons whose effective dimensionality is lower than the effective dimensionality of the surrounding disc.

The conclusion for the present paper is therefore negative but robust: realistic spiral structure does not significantly alter any of the results of Sections~\ref{sec:uniform} and~\ref{sec:ghz}. However, the analysis assumes that the occurrence fraction is independent of spiral structure. Since spiral arms enhance both star formation and supernova activity, the corresponding variation of $f_{\mathrm{SN}}$ could modify that conclusion. That problem, together with the role of corotation and quasi-one-dimensional percolation, remains for future work.

\subsection{Limits of the model}
\label{sec:disc:limits}

Five approximations are common to all results presented here.

First, the Galaxy has been treated as two-dimensional. For an exponential vertical profile with scale height $h_z$, the corresponding correction to the contactability is
\begin{equation}
  \frac{\eta_{3\mathrm{D}}}{\eta_{2\mathrm{D}}}
    = 1 - \frac{2h_{z}^{2}}{(cL)^{2}}
      \left[ 1 - e^{-cL/h_{z}} \left( 1 + \frac{cL}{h_{z}} \right) \right] ,
  \label{eq:thick}
\end{equation}
which approaches unity as $(h_z/cL)^2$. For the thin-disc scale height $h_z\simeq0.3$\,kpc \citep{BlandHawthornGerhard2016}, the correction is approximately $5\%$ at the reference lifetime, but becomes increasingly important for $L\lesssim3\times10^3$ yr. Because the correction rescales $\eta$ uniformly, the correction affects threshold values but not frontier locations.
Dimensionality enters a second time, and less benignly, through the frontier width itself: the confinement expressed by Equation~\eqref{eq:w_scaling} is a two-dimensional result, and in three dimensions the frontier is no longer held within the critical region in the same way \citep{Rosso1986}.
Any treatment in which a third dimension becomes essential --- a thick disc, or a description in which time is carried explicitly --- would therefore have to revisit the width scaling rather than simply rescale the width.

Second, the model neglects the Galactic bar and bulge, treating the disc as a pure exponential profile. This approximation is least important for the Middle and Outer zones, whose frontier structure lies well outside the central kiloparsecs, but the approximation should be borne in mind when interpreting the Inner-zone results.

Third, the occurrence fraction is represented as the product of three independent factors evaluated at the present epoch. Processes such as radial migration would tend to mix stellar populations and further flatten $\zeta(r)$ \citep{Vukotic2016}, reinforcing rather than weakening the principal results of Section~\ref{sec:ghz:field}. The truncation at $R_{\mathrm{tr}}=20$\,kpc removes only $0.4\%$ of the disc and is negligible.

Fourth, and most consequentially, the link rule assumes omnidirectional communication with a fixed range $cL$, so that the communicative lifetime of a civilization also fixes the greatest distance the signals of that civilization reach.
The network is therefore built from a static snapshot in which all civilizations are assumed to coexist, and the network already includes relaying through intermediate civilizations but not time-dependent, directional, or civilization-dependent transmission strategies.
The snapshot overestimates the long-range connectivity that contributes most strongly to the giant component.
The isolation probability draws most of its weight from partners much closer than $cL$ and is far less affected by the snapshot assumption, so the local results of Section~\ref{sec:ghz:solar} are more robust than the giant-component results.
Paper~III relaxes both restrictions: the snapshot is replaced by a spacetime formulation in which a link requires both spatial proximity and temporal overlap of the corresponding light cones \citep{Balbi2018,Grimaldi2018}, and the detection range is separated from the light horizon, so that the two lengths conflated here become independent and the ratio of the two, rather than the communicative lifetime alone, controls the outcome.
Paper~III then determines how much of the structure found here survives.

Fifth, all civilizations are assigned the same communicative lifetime $L$, a single-lifetime idealization equivalent to a delta-function population distribution. The actual distribution is unknown, and \citet{BalbiCirkovic2021} have shown that a long-lived tail in the distribution of technosignature durations can substantially affect detection estimates. The link rule adopted here ties the communication range to the lifetime through $cL$, so a distribution of lifetimes would also be a distribution of ranges. Paper~IV treats that generalization quantitatively, replacing the pair $(L, d_{\max})$ by a joint distribution $P(L, d_{\max})$ within the spacetime framework of Paper~III while holding the mean lifetime fixed. Where the relevant moments are finite, the contactability then depends on the second moment of the lifetime distribution as well as on the mean, the contact counts depend on the covariance between lifetime and reachable area, and the connectivity thresholds depend on further properties of the joint distribution. A spread in lifetime at fixed communication range leaves the Galactic frontiers almost unchanged; it is the correlation between lifetime and range, which the link rule adopted here makes perfect, that carries the effect. What survives unaltered is the shape of the civilization density, so that $r_{\mathrm{peak}}$ and the ratio $\eta_{\max}/\eta(R_{0})$ remain properties of the Galactic density field alone. The frontier positions and widths do not: in the correlated populations examined in Paper~IV the frontier is both displaced outward, by up to $4$\,kpc, and broadened, from about $1.4$ to $2.4$ scale lengths, because a distribution of ranges superposes frontiers at different radii, while the non-detection bound of Section~\ref{sec:uniform:eta0} tightens from $\eta(R_{0}) \le 2.75$ to approximately $0.9$. That broadening acts in the same direction as the principal conclusion of this paper rather than against it. The frontier positions, widths and lifetime constraints reported here should therefore be read as conditional on the single-lifetime snapshot model rather than as quantities invariant under an arbitrary lifetime distribution.

Two parameters carry the remaining uncertainty, and the two parameters are of very unequal weight.
The location of the GHZ is the dominant parameter: because $r_{\mathrm{peak}} = R_{0} + h_{R} \ln k_{\mathrm{SN}}$, that location is controlled almost entirely by the supernova sterilization parameter, and constraining the survival fraction $f_{\rm SN}(R_{0}) = e^{-k_{\rm SN}}$ is therefore the key observational problem for distinguishing between the Inner, Middle and Outer realizations considered here.
The Drake parameter is much less influential: since $\eta \propto D L^{3}$, frontier locations vary only as $D^{-1/3}$ and frontier widths only as $D^{-1/7}$, so the structure identified in this paper is set by the Galactic density field and the connectivity thresholds themselves rather than by the precise value of $D$.


\section{Conclusions} \label{sec:conclusions}

We have extended the uniform-disc framework of Paper~I to a Galaxy with an exponential stellar disc and a radially varying occurrence fraction. The resulting Galactic civilization networks are governed by the local contactability $\eta(R_0)$, which is constrained by our own non-detection.

\begin{enumerate}

\item The contact and network frontiers are intrinsically broad, spanning $3.08\,h_R$ and $1.50\,h_R$ respectively, and the width of the network frontier scales only weakly with the Drake parameter, as $D^{-1/7}$. The two frontiers are separated by $\Delta r = h_R \ln \eta_{\rm network}$, yet the combined half-widths exceed that separation, so the two transition bands overlap and the Galaxy acquires contact and loses network connectivity over the same radial interval. Within the Galactic density models considered here, the connectivity transition therefore does not form a sharp radial boundary.

\item A GHZ produces an annular civilization distribution whose peak radius is
$r_{\mathrm{peak}} = R_{0} + h_{R} \ln k_{\mathrm{SN}}$.
The wide range of published GHZ locations can therefore be understood primarily as differences in the assumed strength of supernova sterilization.

\item The position of the GHZ has only a modest effect on our own prospects for contact, but the position of the GHZ strongly influences the connectivity of the rest of the Galaxy.

\item The key quantity controlling Galactic connectivity is $\eta_{\max}/\eta(R_0)$, a ratio that is itself independent of $\eta(R_0)$. Connectivity is minimized when the GHZ is centred near the solar Galactocentric radius, because no region of the Galaxy is then substantially more contactable than the solar neighbourhood. Whether that minimum is low enough to suppress the giant component altogether does depend on $\eta(R_0)$, and the suppression holds across the observationally relevant range $\eta(R_0) \lesssim 3$.

\item Applied locally, the criterion of Paper~I remains accurate for local quantities such as isolation probability, but not for global quantities such as giant-component membership. Paper~I therefore provides a local description of connectivity, whereas the present work supplies the radial structure that this local description averages over.

\end{enumerate}

Taken together, these results describe a Galaxy whose connectivity changes gradually with radius rather than at a boundary, so that the question of whether humanity lies inside or outside an inhabited region has no unique answer; the quantity most directly constrained by observation remains the local contactability $\eta(R_0)$.


\section*{Acknowledgements}
The author used Claude (Anthropic), a large language model, to assist with the development of the manuscript, 
including the drafting and revision of text, improvement of English expression, refinement of argument structure, and preparation of figure concepts. 
The author also used Microsoft 365 Copilot (Microsoft Corporation), a large language model, to obtain a critical review of the manuscript, to perform independent checks of the results, and to assist with the drafting of the corresponding revised passages.
All content generated with AI assistance was critically reviewed, edited, and verified by the author. 
The author assumes full responsibility for the content of the manuscript, including its analyses, interpretations, and conclusions.


\section*{Data Availability}
 
The code and data supporting this study are available from the author upon reasonable request.



\bibliographystyle{mnras}
\bibliography{GCNT2_Tsumura} 








\bsp	
\label{lastpage}
\end{document}